\documentclass[epj]{svjour}          

\RequirePackage[T1]{fontenc}

\smartqed  

\usepackage{amsmath, amssymb, bm, graphicx, graphics, color, mathrsfs}
\RequirePackage{graphicx}
\RequirePackage{mathptmx}      
\RequirePackage{flushend}
\RequirePackage[numbers,sort&compress]{natbib}
\RequirePackage[colorlinks,citecolor=blue,urlcolor=blue,linkcolor=blue]{hyperref}

\journalname{Eur. Phys. J. C}

\newtheorem{prop}{Proposition}
\begin{document}

\title{On stability of a neutron star system in Palatini gravity}
\author{Aneta Wojnar
\thanks{\emph{e-mail:} aneta.wojnar@poczta.umcs.lublin.pl}%
}                     
\offprints{}          
\institute{Institute of Physics,
Maria Curie-Sk{\l}odowska University \\
20-031 Lublin, pl. Marii Curie-Sk{\l}odowskiej 1, Poland\label{addr1}}
\date{Received: date / Revised version: date}
%
\abstract{
We formulate the generalized Tolman-Oppenheimer-Volkoff equations for the $f(\hat{R})$ Palatini gravity
in the case of static and spherical symmetric geometry. We also show that a neutron star can be a stable system independently of
the form of the functional $f(\hat{R})$.
}
\maketitle
\date{Received: date / Accepted: date}
\newcommand  {\Rbar} {{\mbox{\rm$\mbox{I}\+!\mbox{R}$}}}
\newcommand  {\Hbar} {{\mbox{\rm$\mbox{I}\!\mbox{H}$}}}
\newcommand {\Cbar}{\mathord{\setlength{\unitlength}{1em}
     \begin{picture}(0.6,0.7)(-0.1,0) \put(-0.1,0){\rm C}
        \thicklines \put(0.2,0.05){\line(0,1){0.55}}\end {picture}}}
\newcommand{\be}{\begin{equation}}
\newcommand{\ee}{\end{equation}}
\newcommand{\ben}{\begin{eqnarray}}
\newcommand{\een}{\end{eqnarray}}
\section{Introduction}
\label{sec:introduction}
Einstein by formulating his General Relativity \cite{Einstein:1915ca,Einstein:1916vd} showed that there is a relation between geometry of spacetime 
and matter contribution. Until now, many 
astronomical observations have tested GR giving us that it is the best matching theory for explaining gravitational phenomena which we have in hand.
However, there are still many issues in fundamental physics, astrophysics and cosmology
(for review, see e.g. \cite{Copeland:2006wr,Nojiri:2006ri,Capozziello:2007ec,Carroll:2004de,Sotiriou:2008ve}) indicating
 that GR is not a final theory of gravitational interactions. There are such unsolved
problems as the dark matter puzzle \cite{Cap_Laur, Cap_Far}, inflation \cite{Starobinsky:1980te,Guth:1980zm}, another one is the late-time cosmic acceleration \cite{Huterer:1998qv, Sami} which is explained by the assumption that there
exists an exotic fluid called dark 
energy \cite{Copeland:2006wr,Nojiri:2006ri,Capozziello:2007ec}. It is introduced to the dynamics throughout adding the 
cosmological constant to the standard Einstein's field equations considered in Friedmann-Robertson-Lemaitre-Walker (FRLW) 
spacetime.
Due to the above shortcomings one looks for different approaches in order to find a good theory which will be able to answer the above
problems and to show us directions of future research. The additional argument in a favor of searching generalizations of 
gravity is non-renormalization of Einstein's theory. The renormalization problem seems to be solved if extra high curvature 
terms are added \cite{stelle}. One would also like to unify gravity theory with the other ones (electromagnetism, weak and strong
interactions has been already unified into the Standard Model) but no satisfactory result has been obtained so far as e.g. string theory, 
supersymmetry, that could combine particle physics and gravitation.

One of many ways to deal with the mentioned problems are Extended Theories of Gravity (ETG) \cite{Cap_beyond, cap_invar} which
have been gained a lot of attention. The main arguments which advocate them are a possibility to explain problematic issues without introducing exotic, 
immeasurable entities as well as some approaches to quantum gravity \cite{buch} or the geometrization of the Standard Model (plus gravity) as 
derived from Chamseddine-Connes spectral action
in noncommutative geometry \cite{nonc, nonc2} which indicate that the effective action for the classical gravity should 
be more complicated than the one introduced by Einstein. 

The geometric part of the action might be changed in many different ways. One may assume that constant of Nature are not
really constant values \cite{Dabrowski:2012eb, Leszczynska:2014xba, Salzano:2016pny}. A scalar field might be added into Lagrangian 
and moreover, it can be minimally or non-minimally coupled to gravity \cite{brans, Bergmann}. One proposes
much more complicated functionals than the simple linear one used in GR, for example $f(R)$
gravity \cite{buchdahl, Starobinsky:1980te}. The latter approach has gained a lot of interest recently as the extra geometric terms could explain not only dark matter
issue \cite{cap_jcap, cap_not} but also dark energy problem because it produces the accelerated late-time effect at low cosmic densities. The field
equations also differ from the Einstein's ones so they could provide different
behavior of the early Universe. 

The $f(R)$ gravity is usually treated in two different ways: in the metric approach
\cite{Sotiriou:2008rp,Carroll:2004de,Sotiriou:2008ve,DeFelice:2010aj,Will:1993te}
and Palatini one \cite{Palatini:1919di,Sotiriou:2008rp,Capozziello:2011et,Ferraris:1992dx}. The former arises to the fourth 
order differential equations which are more difficult to handle than the second-order field equations of GR
\footnote{ The theory possesses the scalar-tensor representation, that is, it is mathematically equivalent to the Brans-Dicke gravity (with the parameter $\omega=0$) 
with a potential of the scalar field. In this representation one deals with the differential equations of the second order on the metric components and a modified Klein-Gordon 
equation for the scalar field, see the details e.g. \cite{sotir_pal}.}. Another point is that one believes that physical equations of motion should be of
the second order. In contrast to the metric formalism, the Palatini $f(R)$ gravity provides second order differential equations since 
the connection and the metric are treated as independent objects. The Riemann and Ricci tensors are constructed with the connection while
for building the Ricci scalar we also use the physical metric in order to contract the indices. The Palatini approach allows to get
the modified Friedmann equation \cite{alle_bor1, alle_bor2, alle_bor3} in a form that might be compared with the observational
data \cite{bor_kam, BSSW1, BSSW2, BSS, SSB}. It shows the potential of the Palatini formulation when it is applied to the gravitational 
problems \cite{lavinia, roshan}.

There also exist disadvantages of such an approach: being in conflict with the Standard Model of particle physics 
\cite{sotir_pal, Flanagan:2003rb, Iglesias:2007nv, Olmo:2008ye}, 
the algebraic dependence of the post-Newtonian
metric on the density \cite{Olmo:2005zr, Sotiriou:2005xe}, and the complications with the initial values problem in the 
presence of matter \cite{Ferraris:1992dx, Sotiriou:2006hs}, although 
the problem was already solved in \cite{olmo_sanch}. Another one happens at microscopic scales, that is, the theory produces 
instabilities in 
atoms which disintegrate them. One should also mention limitations which arrive  when one treats the extra terms as fluid-like
\cite{zaeem, zaeem2, zaeem3}. However, it was shown \cite{olmo_tri} that high curvature corrections do not cause such problem. 
What is also very promising, some of the Palatini Lagrangians avoid the Big Bang singularity. 
Moreover, the effective dynamics of Loop Quantum
Gravity can be reproduced by the Palatini theory which gives the link to one of approaches to Quantum Gravity \cite{olmo_singh}.
High curvature correction of the form  changes the notion of the independent connection: in the simple Palatini $f(R)$ gravity the
connection is auxiliary field while in the more general Palatini theory it is dynamical without making the equations of motion 
second order in the fields.

The problem which is our interest in this work concerns astrophysical aspects of Palatini gravity, for example black 
holes \cite{diego1,diego2,diego3,diego4}, wormholes \cite{worm} and neutron stars \cite{anab}.
We would like to examine the last objects in the context of Palatini gravity. In 
general, the maximal mass of neutron stars is still an open problem: Einstein's gravity gives us, together with the recent observations, the limit
as $2$M$_\odot$. The pulsar PSR J1614-2230 is found to have the limit $1.97$M$_\odot$ \cite{demorest} while another one 
is Vela X-1 with the mass $\sim 1.8$M$_\odot$ \cite{rawls}. One also supposes the existence more massive neutron stars as for instance the ones with
masses around $2.1$M$_\odot$ \cite{MassNs} and $2.4$M$_\odot$ (B1957+20) \cite{van}. For the recent neutron stars mass determination see \cite{stellarcollapse}.
There also exists "hyperon puzzle": it turns out that equations of state including hyperons make the maximal mass limit in the case of 
neutron stars without magnetic field lower than $2$M$_\odot$ \cite{hyperon1, hyperon2, hyperon3}. Due to that fact it seems that these equations of state could 
not be feasible for massive neutron stars in the framework of General Relativity. Some solutions were already proposed, like for example
hyperon-vector coupling, chiral quark-meson coupling and the existence of strong magnetic 
fields inside the star. An emissary for the latter idea are works indicating an influence of magnetic fluid which increases masses of the stars
\cite{mag1, mag2, mag3}. This arises to a question if neutron stars without strong magnetic field with mass larger than $2$ M$_{\odot}$ can exist
\cite{astash, astash2, astash3} whose structure can be still described by the equations derived from General Relativity.

Moreover, because of the very recent neutron stars' merger observation \cite{gw} the confrontation of gravitational theories with the observational data is an 
appealing task. Neutron stars are perfect objects for tests at high density regimes thus one may perceive possible deviations from GR. It also seems that
using General Relativity as a description of strong gravitational fields \cite{astash3, eksi} can be an extrapolation for fields sourced by neutron 
stars since they are many orders of magnitude larger than the one we may probe by solar system tests. Therefore, likely GR should be modified in a strong gravity regime and 
in the case of large spacetime curvature \cite{berti}.

Since we are interested in Palatini gravity, one should mention another problem which arose throughout the studies on neutron stars in this formalism. 
In \cite{Barausse:2007ys, barau} it was shown that there exist surface singularities of static spherically symmetric objects in the case of polytropic equation 
of state. It would seem that all Palatini $f(\hat{R})$ functionals besides Einstein-Hilbert one could be ruled out since they provide unphysical behavior. However in
\cite{olmo_ns} it was demonstrated that the problem appears because of the particular equation of state rather than the dynamics of the theory itself.
Further approach to the problem \cite{fatibene} indicates that polytropic equation of state is not a fundamental issue but only an approximation of the matter 
content in the star. Nevertheless, they examined the case of problematic matter description according to the Ehlers-Pirani-Schild (EPS) interpretation
\cite{eps, mauro, fatibene1}, which 
we also follow in our work. Considering the conformal metric as the one responsible for the free fall they revealed that in this case the singularities are not 
generated in comparison to metric which was used in \cite{Barausse:2007ys}.

Due to the above brief considerations on neutron stars, there are two main problems concerning these objects: how to model dense matter inside the star, which one 
would like to describe by an equation of state, and viable equations which depict the macroscopic properties, that is, mass and radius. The alternative 
equations, or rather modified Tolman-Oppenheimer-Volkoff equations, describing the macroscopic values of relativistic stars are actively proposed 
\cite{AltNS1, AltNS2, AltNS3, AltNS4, AltNS5, AltNS6, AltNS7} and tested with the observational data. In this work we would also like to challenge this problem 
and hence in the next section we will present briefly $f(\hat{R})$ gravity in Palatini formalism and then we are going to focus on stars' stability problem. We will also provide 
appropriate TOV equations for this generalization of General Relativity. We are using the Weinberg's \cite{weinberg} signature convention, that is, $(-,+,+,+)$.


\section{A brief description of Palatini gravity: frames}
As we have already mentioned, we are interested in Palatini gravity which is a particular case of EPS formalism \cite{eps}. That means that geometry of spacetime 
is described by two structures which in Palatini gravity turns out to be the metric $g$ and the connection $\hat{\Gamma}$ which are independent objects. 
In addition, the connection is a Levi-Civita connection of a metric conformally related to $g$. Due to this interpretation, one considers motion of a mass particle 
appointed by the connection while clocks and distances are ruled by the metric $g$.
Let us briefly, but in a quite detailed way, introduce the basic of the Palatini $f(\hat{R})$ gravity. The action is given by
\begin{equation}
S=S_{\text{g}}+S_{\text{m}}=\frac{1}{2}\int \sqrt{-g}f(\hat{R}) d^4 x+S_{\text{m}},\label{action}
\end{equation}
where $\hat{R}=\hat{R}^{\mu\nu}g_{\mu\nu}$ is the generalized Ricci scalar. The variation of (\ref{action}) with respect to the metric $g_{\mu\nu}$ provides
\begin{equation}
f'(\hat{R})\hat{R}_{\mu\nu}-\frac{1}{2}f(\hat{R})g_{\mu\nu}=T_{\mu\nu},\label{structural}
\end{equation}
where $T_{\mu\nu}=-\frac{2}{\sqrt{-g}} \frac{\delta}{\delta g_{\mu\nu}}S_m  = (\rho+p)u^{\mu}u^{\nu}+pg^{\mu\nu}$ is energy 
momentum tensor, $p$ pressure while $\rho$ energy density of the fluid. The vector $u^\alpha$ is a $4$-velocity of the observer co-moving with 
the fluid ($g_{\mu\nu}u^\mu u^\nu = -1$).
Taking the trace of the equation (\ref{structural}) with respect to $g_{\mu\nu}$ yields a structural equation, which is given by
\begin{equation}
f'(\hat{R})\hat{R}-2 f(\hat{R})=T.\label{struc}
\end{equation}
Assuming that we are able to solve (\ref{struc}) as $\hat{R}(T)$ we see that $f(\hat{R})$ is also a function
of the trace of the energy momentum tensor, where $T=g^{\mu\nu}T_{\mu\nu}\equiv 3p-\rho$.

The variation with respect to the independent connection gives
\begin{equation}
\hat{\nabla}_\alpha(\sqrt{-g}f'(\hat{R})g^{\mu\nu})=0,\label{con}
\end{equation}
from which we immediately notice that $\hat{\nabla}_\alpha$ is the covariant derivative calculated with respect to $\hat \Gamma$, that is, it is the Levi-Civita connection of the conformal 
metric $h_{\mu\nu}=f'(\hat{R})g_{\mu\nu}.$

 It is well known \cite{DeFelice:2010aj} that the action (\ref{action})
is dynamically equivalent to the  constraint system with first order Palatini gravitational Lagrangian:
\begin{align}\label{action1}
 S(g_{\mu\nu}, \Gamma^\lambda_{\rho\sigma}, \chi)=\frac{1}{2\kappa}\int\mathrm{d}^4x\sqrt{-g}\left(f^\prime(\chi)(\hat R-\chi) + f(\chi) \right)\nonumber\\
 + S_m(g_{\mu\nu},\psi),
\end{align}
providing that $f''(\hat R)\neq 0 $.
\footnote{In that case the linear Einstein-Hilbert Lagrangian $f(\hat R)=\hat R -2\Lambda$ is excluded on that level.}
Introducing further a scalar field $\Phi=f'(\chi)$ and taking into account the constraint equation $\chi=\hat R$, one can rewrite the action in dynamically equivalent way as a Palatini action
\begin{equation}\label{actionP}
 S(g_{\mu\nu}, \Gamma^\lambda_{\rho\sigma},\Phi)=\frac{1}{2k}\int\mathrm{d}^4x\sqrt{-g}\left(\Phi \hat R - U(\Phi) \right)+ S_m(g_{\mu\nu},\psi),
\end{equation}
where the potential $U(\Phi)$ "remembers" the form of function $f(\hat R)$. It is defined as
\begin{equation}\label{PotentialP}
U(\Phi)=\chi(\Phi)\Phi-f(\chi(\Phi))
\end{equation}
where $\chi\equiv\hat R  = \frac{d U(\Phi)}{d\Phi}$
and $\Phi = \frac{d f(\chi)}{d\chi}$.

Palatini variation of this action provides
	\begin{align}
	\label{EOM_P}
	\Phi\left( \hat R_{\mu\nu} - \frac{1}{2} g_{\mu\nu} \hat R \right)   +{1\over 2} g_{\mu\nu} U(\Phi) - \kappa T_{\mu\nu} = 0\\
	\label{EOM_connectP}
	 \hat{\nabla}_\lambda(\sqrt{-g}\Phi g^{\mu\nu})=0\\
	\label{EOM_scalar_field_P}
	  \hat R   -  U^\prime(\Phi) =0
	\end{align}
The last equation due to the constraint $\hat R =\chi= U^\prime(\phi)$ is automatically satisfied. The middle equation (\ref{EOM_connectP}) implies that 
the connection $\hat \Gamma$ is a metric connection for the new metric $h_{\mu\nu}=\Phi g_{\mu\nu}$. The $g$-trace of the first 
equation  $-\Phi \hat R+2 U(\Phi)=\kappa T$, can be recast  as
\begin{equation}\label{struc2}
  2U(\Phi)-U'(\Phi)\Phi=\kappa T.
\end{equation}
which provides an analog of the structure equation (\ref{struc}).
 
Now the equation (\ref{EOM_P}) can be rewritten as a dynamical equation for the 
metric $h_{\mu\nu}$ \cite{BSS,SSB}
 \begin{subequations}
	\begin{align}
	\label{EOM_P1}
	 \bar R_{\mu\nu} - \frac{1}{2} h_{\mu\nu} \bar R  &  =\kappa \bar T_{\mu\nu}-{1\over 2} h_{\mu\nu} \bar U(\Phi)
	\end{align}
	\begin{align}
	\label{EOM_scalar_field_P1}
	  \Phi\bar R &  -  \Big(\Phi^2\,\bar U(\Phi)\Big)^\prime =0
	\end{align}
\end{subequations}
where we have introduced $\bar U(\phi)=U(\phi)/\Phi^2$ and appropriate energy momentum tensor $\bar T_{\mu\nu}=\Phi^{-1}T_{\mu\nu}$.
Moreover, one notices that $\hat R_{\mu\nu}=\bar R_{\mu\nu}, \bar R= h^{\mu\nu}\bar R_{\mu\nu}=\Phi^{-1} \hat R$ and $h_{\mu\nu}\bar R=\ g_{\mu\nu}\hat R$.
The last equation, together with the trace of (\ref{EOM_P1}), can be replaced by
\begin{equation}\label{EOM_P1c}
 \Phi\,\bar U^\prime(\Phi)  + \kappa \bar T = 0\,.
\end{equation}
Thus the system (\ref{EOM_P1}) - (\ref{EOM_P1c}) corresponds to a scalar-tensor action for the metric $h_{\mu\nu }$ and (non-dynamical) scalar field $\Phi$
\begin{equation}\label{action2}
 S(h_{\mu\nu},\Phi)=\frac{1}{2\kappa}\int\mathrm{d}^4x\sqrt{-h}\bigg(\bar R- \bar U(\Phi) \bigg) + S_m(\Phi^{-1}h_{\mu\nu},\psi),
\end{equation}
where
\begin{equation}\label{em_2}
    \bar T^{\mu\nu} =
-\frac{2}{\sqrt{-h}} \frac{\delta}{\delta h_{\mu\nu}}S_m  = (\bar\rho+\bar p)\bar u^{\mu}\bar u^{\nu}+ \bar ph^{\mu\nu}=\Phi^{-3}T^{\mu\nu}~,
\end{equation}
and $\bar u^\mu=\Phi^{-{1\over 2}}u^\mu$, $\bar\rho=\Phi^{-2}\rho,\ \bar p=\Phi^{-2}p$, $\bar T_{\mu\nu}= \Phi^{-1}T_{\mu\nu}, \ \bar T= \Phi^{-2} T$ (see e.g. \cite{DGB}).
Further, the trace of (\ref{EOM_P1}), provides
\begin{equation}\label{EOM_metric_2}
    \bar R= 2\bar U(\Phi)-\kappa \bar T.
\end{equation}


\section{Stability condition}
Following \cite{weinberg} and \cite{green} we wish to show that Palatini neutron stars whose matter is modeled by perfect fluid are in dynamical equilibrium and are
described by modified TOV equations. Moreover, we are going to demonstrate that this equilibrium is stable which means that a considered configuration is in 
thermodynamic equilibrium.
\subsection{TOV equations in ETGs}
It was shown in \cite{aw1} how some modifications of the TOV equations influence a mass and radius of a neutron star. The introduced parameters to the equations not only 
allow to see how the mass-radius diagram changes with respect to them but also can be useful to give some clue on a kind of ETG which we should look for in order
to get the desired results which are in agreement with the observations. The details and discussion on gravitational theories which are related to the 
parameters can be also found in \cite{aw1, szwab}.

The TOV equations can be generalized for a certain class of theories after the generalization of the energy density and pressure \cite{aw2}:
\begin{align}
 Q(r):=\rho(r)+\frac{\sigma(r)W_{tt}(r)}{\kappa B(r)},\\
\label{def2} \Pi(r):=p(r)+\frac{\sigma(r)W_{rr}(r)}{\kappa A(r)}
\end{align}
where the extra terms above are provided by the modified Einstein field equations of the particular form \cite{mim, mim2, mim3}
\begin{equation}\label{mod1}
 \sigma(\Psi^i)(G_{\mu\nu}-W_{\mu\nu})=\kappa T_{\mu\nu}.
\end{equation}
As usually, the tensor $G_{\mu\nu}=R_{\mu\nu}-\frac{1}{2}Rg_{\mu\nu}$ is the Einstein tensor, $\kappa=-8\pi G$, the factor $\sigma(\Psi^i)$ is a coupling to the gravity
while $\Psi^i$ represents for instance curvature invariants or other fields, like
scalar ones. The symmetric tensor $W_{\mu\nu}$ denotes additional geometrical terms which might appear in considered ETG. It should be
noted that (\ref{mod1}) provides a parameterization of gravitational theories at the level of field equations. 
The energy-momentum tensor $T_{\mu\nu}$ will be considered as the one of a perfect fluid discussed in the previous section.
Because the tensor $W_{\mu\nu}$ usually includes extra fields like scalar or electromagnetic ones, apart the modified Einstein's
field equations (\ref{mod1}) we also deal with equations for the additional fields.

As shown \cite{aw1}, in the case of the spherical-symmetric metric
\begin{equation}\label{metric}
 ds^2=-B(r)dt^2+A(r)dr^2+r^2d\theta^2+r^2\sin^2{\theta} d\varphi^2,
\end{equation}
the modified Einstain field equations of the form (\ref{mod1}) reproduces the generalized TOV equations
\begin{align}\label{tov}
  \left(\frac{\Pi}{\sigma}\right)'&=-\frac{G\mathcal{M}}{r^2}\left(\frac{Q}{\sigma}+\frac{\Pi}{\sigma}\right)
  \left(1+\frac{4\pi r^3\frac{\Pi}{\sigma}}{\mathcal{M}}\right)\left(1-\frac{2G\mathcal{M}}{r}\right)^{-1}\nonumber\\ 
	&+\frac{2\sigma}{\kappa r}\left(\frac{W_{\theta\theta}}{r^2}-\frac{W_{rr}}{A}\right)\\
	\mathcal{M}(r)&= \int^r_0 4\pi \tilde{r}^2\frac{Q(\tilde{r})}{\sigma(\tilde{r})} d\tilde{r}.\label{mod_mass}
\end{align}
A specification of a gravitational theory allows us to study stellar configurations. We would like to emphasize the importance of extra field equations which may 
appear in the considered theory of gravity, as we have already shown it in the previous work as well as we will always take them into account further. Moreover, 
let us notice that the modified mass equation (\ref{mod_mass}) comes directly from the solution of the field equations (\ref{mod1}) with the metric
ansatz (\ref{metric}) so the extra term 
proportional to $W_{tt}$ should not be skipped in the examination of the star's configuration. One may treat them as a contribution from additional fields to the 
theory, for example electromagnetic or scalar ones, as caused by the influence of modified equation of state \cite{blas1, blas2, blas3} or as just
pure geometric modifications.

\subsection{Stability of Palatini stars}

Let us focus our attention on the discussed stability problem in the case of Palatini gravity. From the structural equation (\ref{struc}) we expect that the crucial 
quantity in the Palatini formalism, that is, the conformal factor $\phi=f'(\hat{R})$, will depend on the trace of the energy momentum tensor which in the case of the perfect 
fluid has a form $T=3p-\rho$. As we assumed the spherical-symmetric geometry, all physical quantities depend on the radial coordinate $r$ only and hence $\phi=\phi(r)$
which makes the both metric spherical-symmetric.
Another important observation is that the "Palatini-Einstein" tensor $\hat{G}_{\mu\nu}=\hat{R}_{\mu\nu}-\frac{1}{2}g_{\mu\nu}\hat{R}$ built of both, Palatini-Ricci 
tensor and the metric $g_{\mu\nu}$ turns out to be equal to the Einstein tensor considered in so-called Einstein
frame $\bar{G}_{\mu\nu}=\bar{R}_{\mu\nu}-\frac{1}{2}h_{\mu\nu}\bar{R}$. Due to that fact the equations in both frames give a solution on the metric $h_{\mu\nu}$ while 
the metric $g_{\mu\nu}$ can be obtained from the conformal relation (\ref{con}). Therefore we will refer to the metric (\ref{metric}) as the metric $h_{\mu\nu}$ which 
enters the Einstein equations in the Einstein frame (\ref{EOM_P1}). This is a metric responsible for a geodesic motion, that is, it is a Levi-Civita metric of the 
connection $\hat{\nabla}_\mu$. According to this interpretation \cite{eps} motion of particles should be considered with respect to the 
Palatini connection while the metric $g$ is related to light cones and distances which are measured by a normalized observer $u^\mu$, that is, $u^\mu u^\nu g_{\mu\nu}=-1$.

Writing the equations (\ref{EOM_P1}) as 
\begin{equation}
\bar{R}_{\mu\nu}=\kappa\bar{T}_{\mu\nu}-\frac{1}{2}h_{\mu\nu}(\bar{T}+\bar{U}(\phi))
\end{equation}
and using the spherical-symmetric ansatz on the metric $h_{\mu\nu}$ one may write
 \begin{equation}
 \frac{\bar{R}_{rr}}{2A}+\frac{\bar{R}_{00}}{2B}+\frac{\bar{R}_{\theta\theta}}{r^2}=-\frac{A'}{rA^2}-\frac{1}{r^2}+\frac{1}{Ar^2}=\kappa\bar{\rho}+\frac{1}{2}r^2\bar{U}
\end{equation}
from which immediately we get that
\begin{equation}\label{mod_geo}
 A(r)=\left( 1-\frac{2G \mathcal{M}(r)}{r} \right)^{-1},
\end{equation}
where
\begin{equation}\label{nowem}
\mathcal{M}(r)=\int^r_0\left( 4\pi \tilde{r}^2\bar{\rho}(\tilde{r})+\frac{\tilde{r}^2\bar{U}(\tilde{r})}{4G} \right)d\tilde{r}
\end{equation}
or, if one prefers to write the mass distribution for the physical energy density
\begin{equation}\label{nowem2}
 \mathcal{M}(r)=\int^r_0\left( 4\pi \tilde{r}^2\frac{\rho}{\phi(\tilde{r})^2}(\tilde{r})+\frac{\tilde{r}^2U(\tilde{r})}{4G\phi(\tilde{r})^2} \right)d\tilde{r}.
\end{equation}
Let us underline that we took into account the transformations of (\ref{em_2}) together with the conformal transformation of the metric, that is, the diagonal element of the metric $g$ 
will be denoted by $\tilde{A}(r)=\phi^{-1}A(r)$ and so on, which will be also transparent in the further considerations.

Due to the relations (\ref{def2}) and the equations (\ref{mod1}) one has
\begin{subequations}\label{defq}
 \begin{equation}
   \bar{Q}=\bar{\rho}+\frac{1}{2}\bar{U}=\frac{\rho}{\phi^2}+\frac{U}{2\phi^2}=\phi^{-2}Q,\label{q}
 \end{equation}
\begin{equation}
  \bar{\Pi}=\bar{p}-\frac{1}{2}\bar{U}=\frac{p}{\phi^2}-\frac{U}{2\phi^2}=\phi^{-2}\Pi\label{pi}.
\end{equation}
\end{subequations}
We are dealing with two conserved quantities in Einstein frame
\begin{align}
 \hat{\nabla}_\mu\bar{J}^\mu=\hat{\nabla}_\mu n\bar{u}^\mu&=0,\label{current}\\
 \bar{u}_\nu\hat{\nabla}_\mu\bar{T}_\text{eff}^{\mu\nu}=\bar{u}_\nu\hat{\nabla}_\mu \left(\bar{T}^{\mu\nu}-\frac{1}{2}h^{\mu\nu}\bar{U}\right)&=0\label{bian},
\end{align}
where the first equation is the nucleon number current conservation law while the second one is the Bianchi identity contracted with the vector 
field $\bar{u}^\alpha$, which naively could be thought as an observer with respect to the metric $h$ but she is not the one responsible for the measurements. Since the relation 
between the physical observer and the vector field $\bar{u}^\alpha$ is as already mentioned
\begin{equation}
      \bar{u}^\alpha=\phi^{-\frac{1}{2}}u^\alpha                                                                 
\end{equation}
we may apply the vector field $\bar{u}^\alpha$ to the (\ref{bian}) in order to simplify the calculations. The another observation is that the proper nucleon number density $n$
appearing in (\ref{current}) is 
frame independent, that is, $n=-u_\alpha J^\alpha_N=-\bar{u}_\alpha \bar{J}^\alpha_N$. From (\ref{bian})
\begin{align}
 \bar{u}^\mu\left( -n\bar{p}\hat{\nabla}_\mu\Big(\frac{1}{n}\Big)-n\hat{\nabla}_\mu\Big(\frac{\bar{\rho}}{n}\Big)-\frac{1}{2\kappa}\Bar{U}'\hat{\nabla}_\mu\phi \right)=0
\end{align}
which we write as
\begin{equation}
 \hat{\nabla}_\mu n=\frac{n}{\bar{\rho}+\bar{p}}\hat{\nabla}_\mu\bar{\rho}+\frac{1}{2\kappa}\frac{n}{\bar{\rho}+\bar{p}}\bar{U}'\hat{\nabla}_\mu\phi,
\end{equation}
where we may also use the relation (\ref{EOM_P1c}) in the last term.
Thus the infinitesimal changes with respect to the infinitesimal chan\-ges of the energy density are
\begin{equation}
 \delta n=\frac{n}{\bar{\rho}+\bar{p}}\delta\bar{\rho}+\frac{1}{2\kappa}\frac{n}{\bar{\rho}+\bar{p}}\bar{U}'\delta\phi,
\end{equation}
which turns out to be 
\begin{equation}\label{deltan}
 \delta n=\frac{n}{\bar{Q}+\bar{\Pi}}\delta\bar{Q}
\end{equation}
after applying the definitions (\ref{defq}) and $\delta\bar{\rho}=\delta\bar{Q}-\frac{\bar{U}'}{2\kappa}\delta\phi$.

In order to examine the stability condition in the case of Palatini gravity, we will follow the approach presented, for example, in \cite{weinberg}, that is, the 
Lagrange multipliers method. 
Following Weinberg, let us formulate the proposition:
\begin{prop}
 A particular stellar configuration in the Palatini gravity with an arbitrary function $f(\hat R)$, with uniform entropy per nucleon and chemical composition, 
 will sa\-tisfy the equations ($\phi=f'(\hat{R})$)
 \begin{align*}
\mathcal{M}(r)&=\int^r_0 4\pi \tilde{r}^2\frac{Q(\tilde{r})}{\phi(\tilde{r})^2}d\tilde{r},\\
   \frac{d}{dr}\Big( \frac{\Pi(r)}{\phi^2(r)}\Big)&=-\frac{\tilde{A}G\mathcal{M}}{r^2}\left(\frac{\Pi+Q}{\phi({r})^2}\right)\left(1+4\pi r^3\frac{\Pi}{\phi({r})^2\mathcal{M}}\right)
\end{align*}
for equilibrium, if the quantity $\mathcal{M}$, defined by
$$
\mathcal{M}\equiv\int 4\pi r^2\frac{Q(r)}{\phi({r})^2}dr
$$
is stationary with respect to all variations of $\bar{Q}(r)$ that leave unchanged the quantity (nuclear number)
$$
  N=\int^R_0 4\pi r^2\left(1-\frac{2G\mathcal{M}(r)}{r}\right)^{-\frac{1}{2}} n(r)dr
$$
and that leave the entropy per nucleon and the chemical composition uniform and unchanged. The equilibrium is stable with respect to radial oscillations if
$\mathcal{M}$ is a minimum with respect to all such variations.
\end{prop}
In order to demonstrate the validity of the proposition, let us firstly notice that the nuclear number $N$ also depends on modified geometry
via the relation (\ref{nowem}) or (\ref{nowem2}). Moreover, the extra term appearing in (\ref{tov}) vanishes in our particular theory. 
Then we see that $\mathcal{M}$ is stationary with 
respect to all variations that leave the nuclear number fixed if there exists a constant $\lambda$ such that $\mathcal{M}-\lambda N$ is stationary with
respect to all variations. It means that we are interested in the case when $\mathcal{M}$ can be an extremum for the unchanged nuclear number, together with uniform 
entropy per nucleon mass and the chemical decomposition.
Hence we deal with
\begin{eqnarray}\label{deltaMlambda}
 &0&=\delta \mathcal{M}-\lambda \delta N=\int_0^\infty 4\pi r^2 \delta \bar{Q} dr \nonumber\\
 &-&\lambda \int_0^\infty 4\pi r^2\left(1-\frac{2G\mathcal{M}(r)}{r}\right)^{-\frac{1}{2}}\delta n(r)dr\nonumber\\
 &-&
 \lambda G\int_0^\infty 4\pi r\left(1-\frac{2G\mathcal{M}(r)}{r}\right)^{-\frac{3}{2}}n(r)\delta\mathcal{M}(r)dr.
\end{eqnarray}
Using the relation (\ref{deltan}) and gravitational mass definition (\ref{nowem})
\begin{equation} 
 \delta\mathcal{M}(r)=\int^r_04\pi r'^2\delta\bar{Q}(r')dr',
\end{equation}
we interchange the $r$ and $r'$ integrals in the last term after replacing it by the above to get
\begin{align}\label{condit}
 \delta\mathcal{M}&-\lambda\delta N=\int^\infty_0 4\pi r^2\left( 
 1-\frac{\lambda n(r)}{\bar{\Pi}(r)+\bar{Q}(r)}\left[ 1-\frac{2G\mathcal{M}(r)}{r} \right]^{-\frac{1}{2}}\right.\nonumber\\
 &\left.-\lambda G\int^\infty_r 4\pi r'n(r')\left[ 1-\frac{2G\mathcal{M}(r')}{r'} \right]^{-\frac{3}{2}}dr'
 \right)\delta\bar{Q}(r)dr.
\end{align}
From the above expression one sees that $\delta\mathcal{M}-\lambda\delta N$ vanishes for all $\delta\bar{Q}(r)$ if and only if
\begin{align}
 \frac{1}{\lambda}&=\frac{n(r)}{\bar{\Pi}+\bar{Q}}\left[ 1-\frac{2G\mathcal{M}(r)}{r} \right]^{-\frac{1}{2}}\nonumber\\
 &+G\int^\infty_r 4\pi r' n(r')\left[ 1-\frac{2G\mathcal{M}(r)}{r} \right]^{-\frac{3}{2}}dr'
\end{align}
for some the Lagrange multiplier $\lambda$. That means that the right-hand side the above result should be independent of $r$ and hence
\begin{align}\label{rach}
 0=&\left( \frac{n'}{\bar{\Pi}+\bar{Q}}-\frac{n(\bar{\Pi}'+\bar{Q}')}{(\bar{\Pi}+\bar{Q})^2} \right)\left[ 1-\frac{2G\mathcal{M}(r)}{r} \right]^{-\frac{1}{2}}\nonumber\\
 &+\frac{Gn}{\bar{\Pi}+\bar{Q}}\left(4\pi r\bar{Q}-\frac{\mathcal{M}}{r^2}\right)\left[ 1-\frac{2G\mathcal{M}(r)}{r} \right]^{-\frac{3}{2}}\nonumber\\
 &-4\pi Grn\left[ 1-\frac{2G\mathcal{M}(r)}{r} \right]^{-\frac{3}{2}}.
\end{align}
Since the entropy per nucleon is uniform, thus using
\begin{equation}
 n'(r)=\frac{n(r)\bar{Q}'(r)}{\bar{\Pi}+\bar{Q}}
\end{equation}
we write (\ref{rach}) in the form of the TOV equations in Palatini gravity with an arbitrary Lagrangian function as
\begin{align}\label{tov_kon}
  \left(\frac{\Pi}{\phi({r})^2}\right)'&=-\frac{G\tilde{A}\mathcal{M}}{r^2}\left(\frac{Q+\Pi}{\phi({r})^2}\right)
  \left(1+\frac{4\pi r^3\frac{\Pi}{\phi({r})^2}}{\mathcal{M}}\right)\nonumber\\ 
	\mathcal{M}(r)&= \int^r_0 4\pi \tilde{r}^2\frac{Q(\tilde{r})}{\phi(\tilde{r})^2} d\tilde{r},
\end{align}
where $\tilde{A}=\phi^{-1}A(r)$.  Thus, $\delta\mathcal{M}$ vanishes for all $\delta Q$ with the constraint $\delta N=0$ if and only if one deals
with (\ref{tov_kon}). We see from (\ref{condit}) that one deals with a stationary point of $\mathcal{M}$ if the 
 expression on the right-hand 
 side vanishes. It will be so if we are equipped with the generalized TOV equations in the given form since this expression can be transformed into it.

 We have just shown that the mass $\mathcal{M}$ appearing in the TOV equations is stationary, so the necessary condition for the star to pass from stability to 
 instability with respect to some radial perturbation is satisfied. In order to have the TOV equations which describe a stable star's configuration, the constrained extremum must be a minimum:
 therefore one deals with $\delta^2\mathcal{M}>0$. We see from the form of (\ref{nowem2}) that the second order in $\delta\bar{Q}$ has to be positive-definite for all 
 perturbations, analogously to the $\delta\rho$ in General Relativity. It has to be so since if any of the squared frequencies $\omega_n^2$ from the all perturbation
 modes takes negative values then the frequency is purely imaginary and the time variation of this mode would grow exponentially leading to an unstable system. 

\section{Conclusions}
We have examined the stability condition of neutron stars in the case of the Palatini gravity with an arbitrary Lagrangian functional $f(\hat{R})$. We immediately 
notice that the obtained result (\ref{tov_kon}) reduces to the General Relativity equations describing equilibrium of the relativistic stars, specifically, when 
$f(\hat{R})=\hat{R}$ since $\phi\equiv f'(\hat{R})=1$, where prime denotes here the derivative with respect to the Palatini scalar $\hat{R}$. Then also the term 
proportional to $U(\phi)$, which appears in the definitions of generalized energy density and pressure, vanishes.

The performed analysis on the stability criterion for Palatini gravity, which is the simplest representation of EPS formulation, shows that a stability condition of 
a neutron star is very similar to the one in General Relativity. Let us notice that in the case of General Relativity 
one gives the final statement if the star's system is stable or not after the examination of the stability conditions with respect to an equation of state.
In Palatini gravity one has to do even more: we see that the generalized energy 
density ${Q}$ depends on $\rho$ and the form of the $f(\hat{R})$ via the potential $U$ and coupling $\phi$. Thus, besides the equation of state we need to choose a model which 
we wish to study in Palatini formulation. We would like to add here that in the case when we are able to solve the master equation to get $\hat{R}=\hat{R}(T)$, the 
generalized energy density appears finally to depend on the equation of state only.

With the all discussed points above, and together with the previous result presented in \cite{aw2}, where we considered scalar-tensor gravity with a minimally coupled scalar field and an 
arbitrary potential, we conclude that there are other gravitational theories besides General Relativity which allow us to consider possible 
stable stars' configurations with respect to radial oscillations. Future and more detailed work along these lines is currently underway.




\section*{Acknowledgements}
AW is grateful to the Mainz Institute for Theoretical Physics (MITP) for its hospitality and its partial support during the completion of this work. 
The work is supported by the NCN grant $DEC-2014/15/B/ST2/00089$. 
The author would also like to thank Andrzej Borowiec and Diego Rubiera-Garcia for their comments and helpful discussions.


\end{document}